\documentclass{aa}

\usepackage{graphicx}
\usepackage[varg]{txfonts}
\usepackage{natbib}
\bibpunct{(}{)}{;}{a}{}{,}
\usepackage{color}
\usepackage{mathrsfs}
\usepackage{amsbsy}
\usepackage{mathtools}
\usepackage{tabularx}
\usepackage{multirow}
\definecolor{comm}{rgb}{0,0.7,0}

\definecolor{old}{rgb}{0.6,0.4,0.4}

\definecolor{new}{rgb}{0.3,0.7,0.7}

\definecolor{done}{rgb}{1,0.5,0}

\definecolor{todo}{rgb}{0.8,0,0}

\definecolor{idea}{rgb}{0.3,0.5,1}

\newcommand{\mesa}[1]{MESA}
\allowdisplaybreaks

\begin{document} 

\title{Blue large-amplitude pulsators formed from the merger of low-mass white dwarfs}

\author{Piotr A. Ko{\l}aczek-Szyma\'nski\inst{1,2}\fnmsep\thanks{This research was supported by the University of Li\`ege under the Special Funds for Research, IPD-STEMA Programme.}
\and Andrzej Pigulski\inst{1}
\and Piotr {\L}ojko\inst{1}
}
\institute{University of Wroc\l aw, Faculty of Physics and Astronomy, Astronomical Institute, ul. Kopernika 11, 51-622 Wroc\l aw, Poland
\email{piotr.kolaczek-szymanski@uwr.edu.pl}
\and
Space sciences, Technologies and Astrophysics Research (STAR) Institute, Universit\'e de Li\`ege, All\'ee du 6 Ao\^ut
19c, B\^at.~B5c, 4000 Li\`ege, Belgium
}

\date{Received July 23, 2024; Accepted September 24, 2024}

\abstract
{Blue large-amplitude pulsators (BLAPs) are a recently discovered group of hot stars pulsating in radial modes. Their origin needs to be explained, and several scenarios for their formation have already been proposed.}
{We investigate whether BLAPs can originate as the product of a merger of two low-mass white dwarfs (WDs) and estimate how many BLAPs can be formed in this evolutionary channel.}
{We used the Modules for Experiments in Stellar Astrophysics (MESA) code to model the merger of three different double extremely low-mass (DELM) WDs and the subsequent evolution of the merger product. We also performed a population synthesis of Galactic DELM WDs using the COSMIC code.}
{We find that BLAPs can be formed from DELM WDs provided that the total mass of the system ranges between 0.32 and 0.7\,M$_\sun$. BLAPs born in this scenario either do not have any thermonuclear fusion at all or show off-centre He burning. The final product evolves to hot subdwarfs and eventually finishes its evolution either as a cooling He WD or a hybrid He/CO WD. The merger products become BLAPs only a few thousand years after coalescence, and it takes them 20\,--\,70 thousand years to pass the BLAP region. We found the instability of the fundamental radial mode to be in fair agreement with observations, but we also observed instability of the radial first overtone. The calculated evolutionary rates of period change can be both positive and negative. From the population synthesis, we found that up to a few hundred BLAPs born in this scenario can exist at present in the Galaxy.}
{Given the estimated number of BLAPs formed in the studied DELM WD merger scenario, there is a good chance to observe BLAPs that originated through this scenario. Since strong magnetic fields can be generated during mergers, this scenario could lead to the formation of magnetic BLAPs. This fits well with the discovery of two likely magnetic BLAPs whose pulsations can be explained in terms of the oblique rotator model.}


\titlerunning{BLAPs from merging DELM WDs}
\authorrunning{Ko{\l}aczek-Szyma\'nski et al.}

\maketitle


\section{Introduction}\label{sect:introduction}
Blue large-amplitude pulsators (BLAPs) were discovered as a new class of variable stars by \cite{2013AcA....63..379P,2017NatAs...1E.166P} in the Optical Gravitational Lensing Experiment (OGLE) project data \citep{2003AcA....53..291U}. About a hundred BLAPs are currently known, most of which were found in the data of the aforementioned project \citep{2023AcA....73....1B,2024arXiv240417715B,2024arXiv240416089P}. BLAPs are radially pulsating stars with effective temperatures in the range of 25\,to 32\,kK, pulsation periods between 7 and 75 minutes, and characteristic light curves with a rapid rise and a slower decline in brightness. The ranges of brightness variations are between 0.1 and 0.45 mag in the $I$-band, which fully justifies the use of `large amplitude' in the class name. In the Hertzsprung-Russell (H-R) diagram, BLAPs are located between hot subdwarfs and hot main-sequence (MS) stars. While their effective temperatures are similar to sdB-type hot subdwarfs, their surface gravities are an order of magnitude smaller than in hot subdwarfs.  The properties of these stars have been summarised recently by \cite{2024arXiv240416089P}.

One of the most interesting issues related to BLAPs is their origin and the role of binarity in their formation. It is known that stars with parameters typical of BLAPs cannot form during the typical evolution of a single star. It is therefore reasonable to assume that all BLAPs have a binary origin \citep{2021MNRAS.507..621B}. Since BLAPs are very rare (the discovery of about 100 BLAPs required the analysis of the photometry of millions of stars), the evolutionary channels that lead to their formation must be quite rare and/or the evolution through the BLAP instability region is rapid. Several scenarios for the formation of BLAPs have been proposed so far. These predict a fairly wide range of BLAP masses, from just under 0.3\,M$_\odot$ to over 1\,M$_\odot$. They also predict different internal structures of these stars. BLAPs could thus be stars with He cores burning hydrogen in a shell \citep{2017NatAs...1E.166P,2018MNRAS.478.3871W,2018MNRAS.481.3810B,2020MNRAS.492..232B}, core He-burning stars \citep{2017NatAs...1E.166P}, or shell He-burning stars \citep{2022A&A...668A.112X}. They may also be surviving companions of type Ia supernovae \citep{2020ApJ...903..100M}. In all of these scenarios, binarity is needed to remove much of the mass of the BLAP progenitor, especially to remove much of its hydrogen envelope. According to some of these scenarios, BLAPs should be at present members of binary systems.

A very interesting scenario for the formation of BLAPs has been proposed by \cite{2023ApJ...959...24Z} in which BLAPs are formed by the merger of a helium white dwarf (WD) and an MS star. Another merger scenario not yet realised with modelling was proposed by \cite{2018arXiv180907451C}. In their scenario, BLAPs are formed as a result of the merger of two extremely low-mass WDs. The attractiveness of the merger scenario comes from the fact that mergers can produce stars with strong magnetic fields. As we show in the accompanying paper \citep{Pigulski+2024}, there are two BLAPs that we find to very likely be magnetic BLAPs.

Observations can also shed light on BLAP formation scenarios by identifying binarity. 
Close binary systems with BLAPs can be discovered through observations of proximity effects or eclipses. So far, however, none such BLAP is known. On the other hand, wide systems with long and intermediate orbital periods can be discovered by observations of the light travel-time effect, as is the case of HD\,133729 \citep{2022A&A...663A..62P} and ZGP-BLAP-01 = TMTS-BLAP-1 \citep{2023NatAs...7..223L}.

Our paper is organised as follows. In Sect.~\ref{sect:binary-wds} we review studies devoted to WDs in binary systems. In the next Sect.~\ref{sect:mesa-general}, we describe the methods and results of modelling the mergers of double extremely low-mass WDs and the subsequent evolution of their products. We also focus on analysing their seismic properties, focusing on radial pulsations. In Sect.~\ref{sect:pop-synth-general}, we perform a population synthesis of Galactic double extremely low-mass WDs to estimate the number of BLAPs resulting from the mergers of these systems and to quantitatively examine the evolutionary scenarios leading to this outcome. The most important results and conclusions from our work are summarised in Sect.~\ref{sect:summary}.

\section{Binary white dwarfs}\label{sect:binary-wds}
As mentioned above, \cite{2018arXiv180907451C} have suggested that BLAPs may be the result of the merger of two extremely low-mass WDs. This idea has never been verified by modelling, however. Extremely low-mass white dwarfs \citep[ELM WDs,][and references therein]{2010ApJ...723.1072B,2022ApJ...933...94B,2012ApJ...751..141K,2020ApJ...894...53K,2023ApJ...950..141K} are helium-core WDs (HeWDs)\footnote{As the nomenclature in this topic can be confusing, we clarify the differences between the terms `ELM WDs', `HeWDs' and `low-mass WDs'. Helium WDs and low-mass WDs can be considered as synonyms, referring to all WDs with masses below $\sim$0.5\,M$_\sun$, which are thought to have helium cores \citep[e.g.][]{2016A&A...595A..35I}. ELM WDs, on the other hand, are a unique subgroup of HeWDs, characterised by the lowest masses, smaller than $\sim$0.3\,M$_\sun$. They are always the products of evolution in interacting binary systems, which is not the case for all HeWDs.} with masses $\lesssim0.3\,{\rm M}_\sun$\footnote{The upper mass limit of ELM WDs varies between authors. It depends on whether we define ELM WDs as (i) HeWDs formed by evolution in a close binary system, (ii) HeWDs that would require a time longer than the Hubble time to form by the evolution of an isolated star, or (iii) HeWDs with sufficiently low mass that they do not experience hydrogen shell flashes during contraction. In this paper, we assume that this limit is $0.35\,{\rm M}_\sun$.} that have never managed to ignite helium in their interiors. Their presence in the Galaxy can be explained on the basis of the evolution of stars in binary systems in which, after the formation of a helium core, the stellar envelope is quickly removed before the onset of the $3\alpha$ reactions \citep[][]{1995MNRAS.275..828M,2013A&A...557A..19A,2017MNRAS.466L..63Z,2018ApJ...858...14S,2019ApJ...871..148L}. They cannot form as a result of the evolution of a single star, because the age of the Universe is simply too short for a low-mass star to evolve into an ELM WD in that time.\footnote{Isolated HeWDs are observed. They most likely originated from the evolution of red giants with dust-driven winds enhanced by their high metallicity \citep{1993ApJ...407..649C,1996ApJ...466..359D,2007ApJ...671..761K}. However, this evolutionary channel does not allow for the formation of HeWDs with masses below $\sim$0.25\,M$_\sun$.} 

Extremely low-mass WDs also occur in so-called double-degenerate binary systems \citep[DDs; i.e.~binary systems in which both components are WDs; e.g.][and references therein]{2001A&A...375..890N,2017MNRAS.470.1894K,2019MNRAS.490.5888L}. Their companions may also be HeWDs or even ELM WDs \citep[e.g.][]{2009ApJ...707L..51M,2022ApJ...933...94B,2023ApJ...950..141K,2024A&A...685A...9A}. Such systems are called double HeWDs (DHeWDs) or double ELM WDs (DELM WDs). Some of the DHeWD and DELM WD systems are so tight that they merge within a few hundred Myr due to the emission of gravitational waves \citep[e.g.][]{2011ApJ...737L..23B,2020ApJ...892L..35B,2019Natur.571..528B}. In general, the mergers of DDs are the source of many phenomena and objects in the Universe such as `normal' and subluminous type Ia supernovae \citep[e.g.][]{1960ApJ...132..565H,2017hsn..book..317T,2023RAA....23h2001L}, AM\,CVn stars \citep[e.g.][]{2018RAA....18....9Z,2018A&A...620A.141R}, extreme helium stars and R\,CrB stars \citep[e.g.][]{2012MNRAS.426L..81Z,2014MNRAS.445..660Z}, isolated helium- or hydrogen-rich hot subdwarfs \citep[e.g.][]{2016MNRAS.463.2756H,2018MNRAS.476.5303S,2021MNRAS.504.2670Y}, and single ultra-massive WDs \citep[e.g.][]{2020NatAs...4..663H,2022MNRAS.512.2972W}. 

Given the similar characteristics of BLAPs and single hot subdwarfs, the idea of \cite{2018arXiv180907451C} that BLAPs may originate as a result of the merger of two ELM WDs seems reasonable. Indeed, some papers investigating the origin of hot subdwarfs from DHeWD mergers present models that intersect the BLAP region in H-R and/or Kiel diagrams \citep[e.g.][]{2000MNRAS.313..671S,2012MNRAS.419..452Z,2018MNRAS.476.5303S,2021MNRAS.504.2670Y}. However, the authors of these papers focus only on the evolutionary stage typical of hot subdwarfs. Although the literature on DD mergers is extremely extensive due to their association with supernovae, we are not aware of any published study that analyses the mergers of DELM WDs in the context of the formation of BLAPs. Given these facts, we investigate in this work whether DELM WD mergers (or more generally, DHeWD mergers) could be one of the potential sources of Galactic BLAPs.

\section{Simulations of DELM WD merger products}\label{sect:mesa-general}
\subsection{Methods}\label{sect:mesa-methods}
\subsubsection{General setup}\label{sect:general-setup}
To model the DELM WD merger and the subsequent evolution of its product, we used the Modules for Experiments in Stellar Astrophysics\footnote{\url{https://docs.mesastar.org/en/latest/}} \citep[MESA version~23.05.1,][]{Paxton2011, Paxton2013, Paxton2015, Paxton2018, Paxton2019, Jermyn2023}, a one-dimensional stellar structure and evolution code compiled using the MESA Software Development Kit for Linux \citep[version~23.7.3,][]{2023zndo..10624843T}. In this and the next section, we describe in detail the applied settings and the values of critical parameters in the MESA code. However, we would like to emphasize that all the MESA \texttt{inlists} we used to obtain the results presented in this study are available on Zenodo at the following link\footnote{\url{https://doi.org/10.5281/zenodo.13863170}}. We consider three merger models, corresponding to three different total masses of the merging DELM WDs, namely 0.32, 0.4, and 0.7\,M$_\sun$. For simplicity, we refer in this paper to these models as models A, B, and C, respectively. Their key features are listed in Table~\ref{tab:models}. The models have been chosen such that models A and C intersect the area of the occurrence of BLAPs in the Kiel diagram at its edges, while model B intersects this area centrally.
\begin{table}
\caption{\label{tab:models}Fundamental properties of the three models of merging DELM WDs analysed in our study.}
\centering
\begin{tabular}{ccc}
\hline\hline
\noalign{\smallskip}
Model&$M_{\rm ini}\,({\rm M}_\sun)^\ast$&$M_{\rm HeWD,1}+M_{\rm HeWD,2}\,({\rm M}_\sun)^{\ast\ast}$\\
\noalign{\smallskip}
\hline
\noalign{\smallskip}
A&1.2&0.16\,$+$\,0.16 (0.32)\\
B&1.5&0.20\,$+$\,0.20 (0.40)\\
C&2.5&0.35\,$+$\,0.35 (0.70)\\
\noalign{\smallskip}
\hline
\end{tabular}
\tablefoot{$^\ast$ Initial mass of the MS star that evolved to RGB to obtain the model of ELM WD. $^{\ast\ast}$ Masses of the two DELM WD components whose mergers are analysed in this study. The mass of the resulting BLAP is given in parentheses.}
\end{table}

We assumed that both components of the DELM WD progenitor system had identical initial chemical compositions and initial uniform rotation rates of 10\% of their critical values. Our models are characterised by the initial mass fraction of hydrogen $X=0.7$ and metallicity in two variants, $Z=0.01$ and $Z=0.02$. When initialising the models prior to MS, we assumed a solar-scaled mixture of elements according to \cite{2009ARA&A..47..481A} and the corresponding opacity tables for this mixture. The microphysics data used in our MESA models are described in Appendix~\ref{appendix:mesa}. We used the convective pre-mixing scheme \citep[][their Sect.~5]{Paxton2019} in combination with the Ledoux criterion to define the boundaries of convective instability. Convective mixing was incorporated into the models via mixing length theory in the formalism provided by \cite{1965ApJ...142..841H}, with the value of the solar-calibrated mixing length parameter $\alpha_{\rm MLT}=1.82$ \citep{2016ApJ...823..102C}. We ignored the effects of semiconvection and thermohaline mixing. The overshooting of the material above (and below, if possible) any convective, H- or He-burning region was treated in the exponential diffusion approximation developed by \cite{2000A&A...360..952H} with an adjustable parameter, $f_{\rm ov}=0.015$. MESA uses the mathematical formalism of \cite{2000ApJ...528..368H} and \cite{2005ApJ...626..350H} to apply structural corrections, perform different types of rotationally induced mixing and `diffusion' of angular momentum between adjacent shells. We have included the following rotational mixing mechanisms in MESA: dynamic shear instability, secular shear instability, Eddington-Sweet circulation, Solberg-H\o iland instability, and Goldreich-Schubert-Fricke instability, all described in detail by \cite{2000ApJ...528..368H}. Mass losses due to the radiation- or dust-driven stellar winds have been calculated according to the prescriptions given by \cite{2001A&A...369..574V} and \cite{1975psae.book..229R}, respectively. For detailed information on all the above processes, their implementation in the MESA code, and our choices of `physical switches', we refer the reader to the series of MESA instrumental papers and to our \texttt{inlists} that accompany this paper.

\subsubsection{Four stages of simulation}\label{sect:mesa-stages}
Each of our models consists of four distinct evolutionary stages, which we integrated separately. During the first stage, we created a chemically homogeneous and fully convective model of a single pre-MS star. When the model was in the immediate vicinity of the zero-age MS (ZAMS), we relaxed it to a uniform rotation with an angular velocity corresponding to 10\% of the critical value. During the subsequent evolution, we allowed the differential rotation to develop. We followed the evolution of the model up to the point on the RGB where a helium core of the desired mass formed inside the red giant. We defined the helium core boundary as the distance from the centre of the red giant at which $X=0.01$, so our models of pre-ELM WDs do not have hydrogen envelopes, although we note that some HeWDs can stably burn hydrogen in their relatively thick envelopes \citep[e.g.][]{2010ApJ...718..441S}.

When the ELM WD progenitor formed inside a red giant, we aborted the first stage of evolution and started to artificially remove the hydrogen-rich envelope using the MESA \texttt{control} named \texttt{relax\_mass\_to\_remove\_H\_env}. For the reasons described below, we used this relaxation procedure together with \texttt{extra\_mass\_retained\_by\_remove\_H\_env\,=\,0} to ensure that the surface of the target ELM WD retains as little hydrogen as possible. As a result, the hydrogen-rich envelope was removed at a constant rate of $10^{-3}\,{\rm M}_\sun\,{\rm yr}^{-1}$, simulating efficient mass transfer between components or envelope ejection as a result of the common envelope (CE) phase. Despite the procedure used, the model still contained a small amount of hydrogen near the surface, which tended to burn off rapidly in a series of flashes. However, this had the disadvantage of being computationally expensive and only delayed the descent of the helium core into the HeWD region, while the products of these nuclear reactions were easy to predict. Since we were only interested in the final product, namely the formation of the ELM WD, immediately after the MESA procedure removed the envelope, we synthetically converted the trace amount of the remaining hydrogen into helium to reduce computational time\footnote{A detailed spectroscopic analysis of selected BLAPs by \cite{2024arXiv240416089P} indicates that most of them are characterised by atmospheres with significantly enhanced helium and metal content. The logarithm of the helium-to-hydrogen ratio reaches $\log(N_{\rm He}/N_{\rm H})=-0.5$, which corresponds to $Y\approx 0.55$. Hydrogen is therefore still present in the atmospheres of the observed BLAPs.}. After the removal of the envelope, we continued the evolution of the helium core until its $\log(L/{\rm L}_\sun)=-2$. At this stage, we did not include diffusion and gravitational settling of elements near the surface, as the subsequent dynamical merger led to intense mixing of the surface layers, effectively removing any chemical composition inhomogeneities. The result of this stage was an ELM WD model on a HeWD cooling sequence.

Having a model of a single ELM WD, we started the third stage, the simulation of the coalescence of the DELM WD. We assumed that the merger involved a pair of ELM WDs with identical mass and chemical composition as in the model obtained in the previous stage.\footnote{In reality, there is always some difference between the masses of the DELM WD components, and the less massive component (with a larger radius) undergoes tidal disruption. However, for simplicity, we modelled the whole phenomenon as occurring in a system with twin ELM WDs. It seems to us that a model with a moderate discrepancy between the masses of the components, although more realistic, would not necessarily be more informative in terms of the results we obtained.} Unfortunately, for MESA, simulating a merger of DHeWD is an extremely numerically demanding task, especially in the initial phase when it is difficult to maintain model convergence. This is due to the rapid heating and subsequent expansion of the He-rich matter deposited on the surface of the relatively cool ELM WD ($T_{\rm eff}\approx10\,000\,$K) at a very fast accretion rate. To mitigate this problem, we started the third stage of our simulation with the accretion of a thin layer of matter from the tidally disrupted ELM WD at an accretion rate of $10^{-7}\,{\rm M}_\sun\,{\rm yr}^{-1}$. This relatively slow rate allowed us to guide the evolution of the model through the most challenging phase of the merger simulation in a numerically stable way. When the total mass of He-rich material accumulated on the surface of the accretor reaches 0.01\,M$_\sun$, we increased the accretion rate to 10$^{-5}$\,M$_\sun\,{\rm yr}^{-1}$. This corresponds to the so-called slow merger model \citep[e.g.][and references therein]{2012MNRAS.419..452Z}. This model assumes that the tidally disrupted ELM WD first forms a disc around the accretor, and then the matter from the disc settles relatively slowly (on a time scale of $\sim$10$^4$\,years) on the surface of the accretor.\footnote{Detailed hydrodynamic simulations provide evidence that this process is more complex. After a tidal disruption of a less massive WD, about half of its material strikes directly the surface of the accretor in the form of a stream, immediately forming a `hot corona' of significant size that prevents further settling of matter on the dynamical time scale. In contrast, the remaining matter forms an almost Keplerian-rotating disc from which the matter is consumed by the accretor on a much longer time scale. For a discussion on this topic see e.g.~\cite{2007MNRAS.380..933Y}, \cite{2009A&A...500.1193L}, and \cite{2014MNRAS.438...14D}. Here we have chosen the slow merger model due to its relatively simple implementation in MESA.} The accretion rate we adopted corresponded to about 30\,--\,60\% of the Eddington rate during the main part of the mass accretion process. We stopped the accretion when the total mass of the accretor was equal to the sum of the component masses of the DELM WD, as indicated in Table~\ref{tab:models}. During the accretion of matter, we disabled all rotation-related effects to avoid problems with the convergence of the model. However, hydrodynamic simulations of DHeWD mergers carried out by, for example, \cite{2006MNRAS.371.1381G} and \cite{2014MNRAS.438...14D} suggest that the post-merger product rotates rapidly, and the radial rotational profile quickly approaches rigid rotation. To account for this important feature, we relaxed our model to a rigid rotation with an angular velocity corresponding to 30\% of its critical value \cite[cf.][especially their Fig.\,4 and Table~B.1]{2014MNRAS.438...14D} at the end of the accretion phase. The same authors also provide evidence that as a result of the dynamical interactions during the merger, only $\sim$10$^{-3}$\,M$_\sun$ are irreducibly ejected from the system in the form of a `tidal tail'. Therefore, we could model the whole phenomenon as perfectly conservative in terms of the total mass of the interacting DELM WD system.

During the final, fourth stage of the simulation, we followed the evolution of the post-merger product, focusing on the point at which it passes through the region of BLAPs in the H-R and Kiel diagrams. Although the resulting star rotated relatively fast, suggesting significant rotational mixing in the subsurface layers, we included diffusion and gravitational settling of elements for completeness. These processes were tracked individually for each isotope, without grouping them into larger categories. We terminated the calculations when the post-merger object was on the WD cooling sequence.

\subsection{Results of simulations with MESA}\label{sect:mesa-results}

\subsubsection{Evolution of merger in H-R and Kiel diagrams}\label{sect:hr-kiel}
\begin{figure*}[!ht]
\centering
\includegraphics[width=0.85\textwidth]{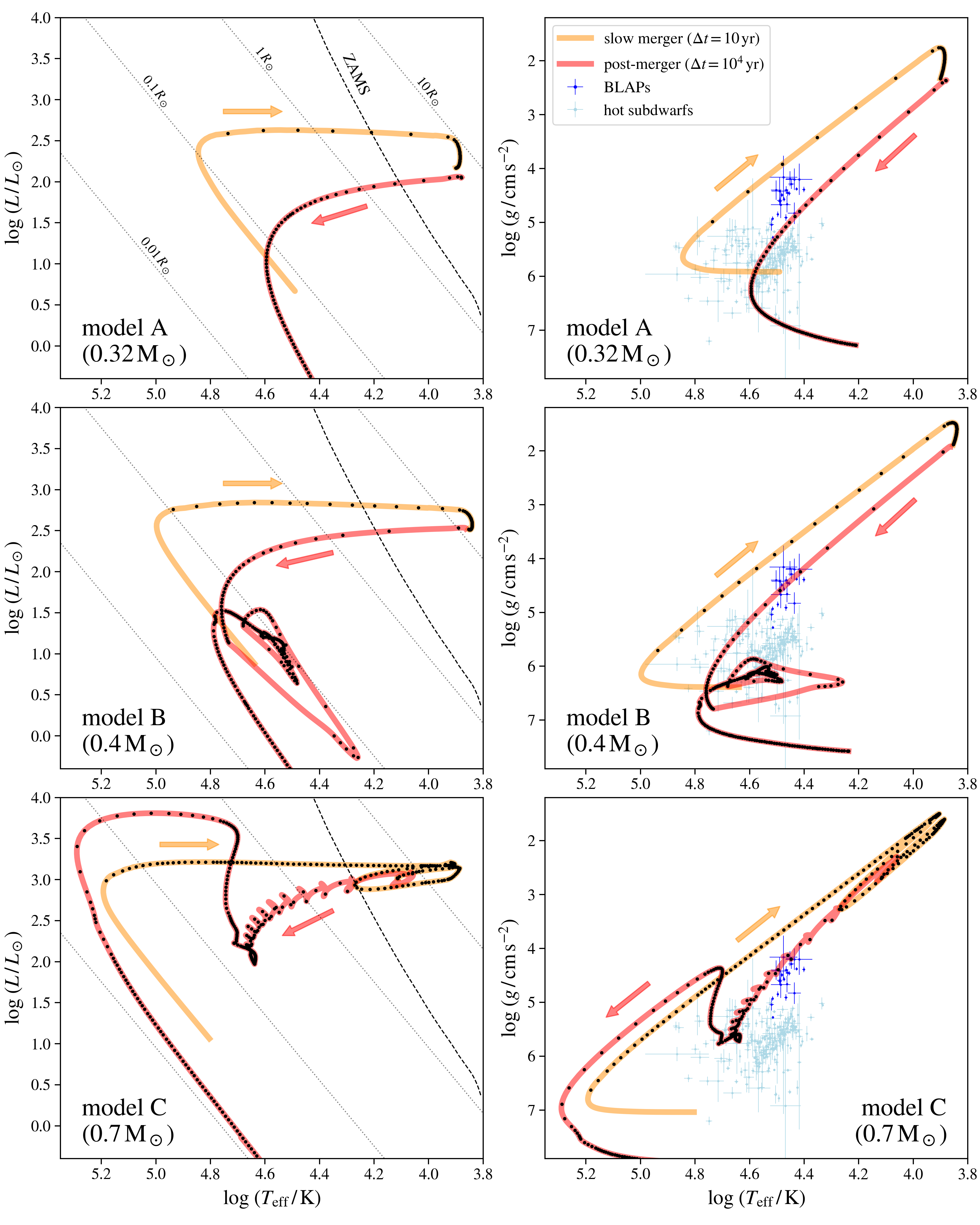}
\caption{Evolutionary tracks for the models A (top row), B (middle row), and C (bottom row) with the initial metallicity $Z=0.02$ in the H-R diagrams (left column) and Kiel diagrams (right column). The orange curves represent the evolution of the model during the slow merger phase, while the red curve corresponds to the evolution of the post-merger product. The black dots along the evolutionary tracks indicate fixed time intervals with the periods provided in the legend. Arrows indicate the direction of evolution at both stages. The dashed curve in the H-R diagrams corresponds to ZAMS. The diagonal dotted lines in the H-R diagrams correspond to the lines of constant radii and are labelled in the top right panel. For comparison, the dark blue and light blue points with error bars correspond to the observed parameters of BLAPs and hot subdwarfs, respectively. Data sources for them are given in the text.}
\label{fig:HR_all_models}
\end{figure*}

Figure~\ref{fig:HR_all_models} illustrates the evolution of the models A, B, and C with initial $Z=0.02$ in the H-R and Kiel diagrams. As can be seen in Fig.~\ref{fig:period_and_eta}, the evolutionary tracks of the merger products with initial $Z=0.01$ are very similar, so we do not plot them in Fig.~\ref{fig:HR_all_models} for greater clarity. We restrict our presentation of the evolutionary tracks to the slow merger\footnote{In Fig.~\ref{fig:HR_all_models} we have omitted the initial merger phase with the accretion rate limited to 10$^{-7}$\,M$_\sun$\,yr$^{-1}$; hence, the grey line does not start at $\log(L/{\rm L}_\sun)=-2$, which characterises the initial state of the ELM WD accretor.} (orange curves) and the post-merger (red curves) phases. The reason why the evolutionary track of the post-merger product does not start exactly at the point in the diagrams where the merger was formed is because the model was relaxed to a uniform rotation after the completion of the merger, as we discussed earlier in the text. For comparison, we have plotted the observed positions of BLAPs \citep{2017NatAs...1E.166P,2024arXiv240416089P,2022MNRAS.513.2215R,2024MNRAS.529.1414C} and hot subdwarfs \citep{2019ApJ...880...79F,2023ApJ...942..109L} in Kiel diagrams.\footnote{We did not draw the positions of the BLAPs on the H-R diagram, because reliably determining their luminosities is a difficult task, as most of them are relatively far away and are characterised by strong interstellar absorption.} We have also plotted the position of the ZAMS in the H-R diagrams according to the MIST evolutionary tracks \citep{2016ApJS..222....8D,2016ApJ...823..102C} taking into account the initial solar metallicity and rotational effects. We describe the behaviour of each model in more detail below.

The evolution during the slow merger phase proceeds similarly in each of the three models. During the first approximately 10 years, the accretor hardly changes its radius, but increases its luminosity by about five orders of magnitude due to a rapid increase in effective temperature. After reaching $\log(L/{\rm L}_\sun)\approx3$, the assumed accretion rate becomes close to the Eddington limit, causing the accretor to maintain a nearly constant luminosity for the next few tens or hundreds of years, while rapidly increasing its radius to approximately $10\,{\rm R}_\sun$. At this stage, the accretor mimics an intermediate-mass MS star with a mass of $\sim$5\,M$_\sun$, although its atmosphere lacks hydrogen. Overall, after $\sim$10$^4$ years, the merger is complete. For the model C, off-centre He burning is initiated during the final accretion phase, causing the model to exhibit a characteristic loop at this stage of the simulation.

Once the merger is completed, the subsequent post-merger evolution differs from model to model. In the model A, the post-merger product simply evolves towards the HeWD cooling sequence, and its interior no longer undergoes thermonuclear reactions. With a mass of 0.32\,M$_\sun$, it marginally `grazes' the region where BLAPs occur. Interestingly, with a mass lower by only 0.02\,M$_\sun$, the evolutionary track of the model A would miss the region of BLAPs. Therefore, we can consider 0.32\,M$_\sun$ as the minimum mass of a BLAP originating from a merger of DELM WD. 

The more massive model B intersects the BLAP region centrally, and its evolution at this phase is similar to the model A. However, a difference appears after a few tens of thousands of years, when He burning begins in the model B. Initially, helium ignites outside the core in a series of flashes, causing the star to make a series of tightening loops in the H-R diagram. The burning then reaches the centre, helium is gradually burned out in the core, and the star becomes a hybrid helium-carbon-oxygen WD (He/CO WD) in which about half the mass is helium. 

The evolution of the model C, with a mass of 0.7\,M$_\sun$, differs from the previous two cases. Off-centre helium burning already starts at the end of the slow merger phase and continues in a series of about ten flashes that alternately increase and decrease the radius of the star. The model C is therefore a BLAP model in which He-burning can occur, in contrast to the models A and B, which are BLAP models without any thermonuclear energy source when crossing the region of BLAPs. Once helium in the core is depleted, the model C also becomes a hybrid He/CO WD with a carbon-oxygen core and a helium envelope, although much thinner than in the model B. Increasing the mass of the model C leads to higher luminosities of the post-merger product as it approaches the BLAP region. Therefore, we consider 0.7\,M$_\sun$ to be an upper limit on the mass of the BLAP formed from the merger of the pair of ELM WDs.

For each model, a post-merger object becomes a BLAP only about four to ten thousand years after the coalescence. This is a relatively short time compared to the later time scale of the evolution of these objects. This allowed us to conclude that a DELM WD can become a BLAP almost immediately after merging. Consequently, it can be expected that BLAPs originating from this evolutionary channel should exhibit some infrared excess emission due to the dust created from the remnants of metal-rich material ejected during the tidal disruption. The time spent in the region occupied by the BLAPs is also relatively short. Regardless of the model, its duration is on the order of tens of thousands of years, which certainly correlates well with the remarkable rarity of BLAPs in the Galaxy. BLAPs formed in the models A, B, and C can be described as pre-hot subdwarfs because they always become hot subdwarfs after leaving the BLAP region. They are also relatively fast rotators. All of our BLAP models exhibit a rotation period of about 0.5\,d and a typical equatorial rotational velocity of around 100\,km\,s$^{-1}$.

\subsubsection{Seismic properties of the BLAP models}\label{sect:seismic-properties}
\begin{figure*}[!ht]
\centering
\includegraphics[width=0.9\textwidth]{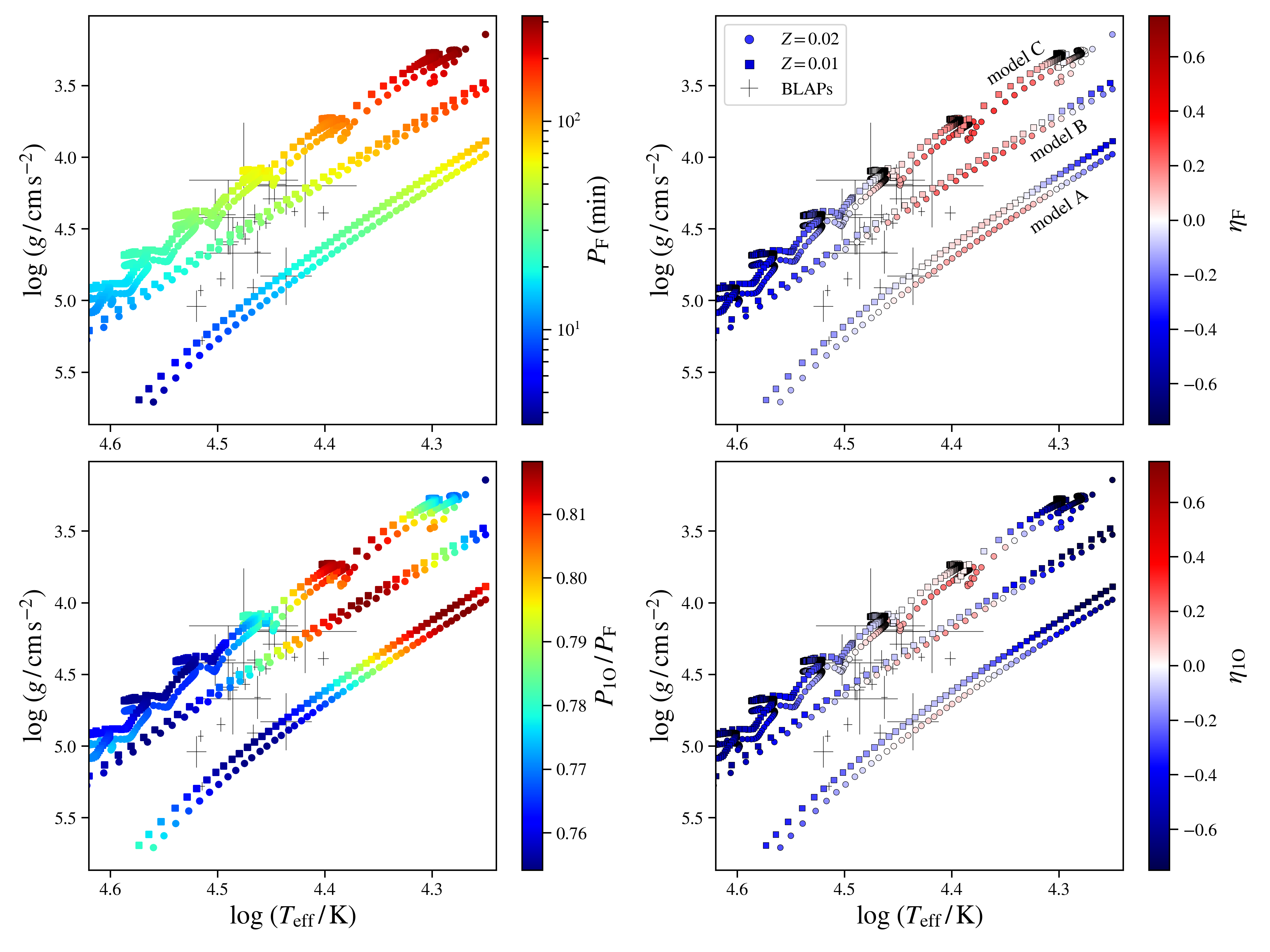}
\caption{Evolution of the seismic properties of the post-merger models A, B, and C (colour-coded points) in the Kiel diagrams as they pass the BLAP region. Colour-coded circles correspond to models with initial metallicity $Z=0.02$, while squares correspond to $Z=0.01$. The black crosses indicate the position of the BLAPs based on the same data as in Fig.~\ref{fig:HR_all_models}. The colour scales in the panels correspond to the pulsation period of the fundamental radial mode (upper left), the normalised growth rate of this mode (upper right), the period ratio of the first radial overtone to the fundamental radial mode (lower left) and the normalised growth rate of the first overtone (lower right).}
\label{fig:period_and_eta}
\end{figure*}

Tracing the evolution of the post-merger product in MESA, we also investigated its seismic properties using the GYRE\footnote{\url{https://gyre.readthedocs.io/en/latest/}} code \citep{2013MNRAS.435.3406T,2018MNRAS.475..879T,2020ApJ...899..116G} for linear non-adiabatic stellar oscillations. As input files for GYRE with the necessary stellar structure data, we used \texttt{profiles} saved by MESA. Since BLAPs are expected to pulsate in radial modes, we are only interested in the period of the radial fundamental mode, $P_{\rm F}$, and the period of its first overtone, $P_{\rm 1O}$. We also investigate the stability of these two modes. Our calculations in GYRE were performed in the non-adiabatic regime with \texttt{MAGNUS\_GL2} difference equation scheme. The corresponding \texttt{gyre.in} file with the input parameters we used can be found in the Zenodo repository, which we mentioned in Sect.~\ref{sect:general-setup}. 

We monitored the evolution of post-merger models with diffusion and gravitational settling of elements incorporated. Nevertheless, it turns out that the rotation rate of the post-merger product is sufficiently fast that, once the BLAP region was crossed, our three models show significant rotational mixing in the outer layers. This process is dominated by the Goldreich-Schubert-Fricke instability and Eddington-Sweet circulation, with a combined diffusion coefficient of $\sim$10$^6$\,cm$^2$\,s$^{-1}$ in the subsurface layers. Such intensive rotational mixing effectively prevents the formation of any chemical inhomogeneities near the surface, with the result that our seismic models of BLAPs have a highly homogeneous chemical composition. 

The results of the seismic analysis of our models are summarised in Fig.~\ref{fig:period_and_eta}. As can easily be seen, the period of fundamental radial mode (upper left panel) changes rapidly at this stage of evolution as a consequence of rapid changes in the radius of the star. Each model correctly reproduces the range of pulsation periods observed in BLAPs (which range from 7 to about 75 minutes, Sect.~\ref{sect:introduction}). Analysis of the stability of the fundamental radial mode (upper right panel) reveals a satisfactory agreement between the location of unstable modes and the observed positions of BLAPs. The exact location of this `area of instability' significantly depends on the choice of opacity tables, initial metallicity, and the structure of the outer layers, where the radial pulsations in BLAPs are driven in the so-called $Z$-bump region. Given the large uncertainties associated with the mean opacities of the iron group elements \citep[see e.g.][and discussion therein]{2017MNRAS.466.2284D}, our result can be considered satisfactory. We also cannot exclude the possibility that a small addition of hydrogen near the surface \citep[which could potentially survive the merger; e.g.][]{2016MNRAS.463.2756H} could change the structure of the outer layers enough to alter the geometrical location of the $Z$-bump, consequently affecting the position of the instability region of the fundamental radial mode that we obtained. We would like to point out that the observational sample of BLAPs does not correspond to a single evolutionary channel discussed in our study. Most likely, BLAPs originate from several different evolutionary channels. For this reason, their internal structure, chemical composition, and rotational profile may differ significantly from object to object. Consequently, our models do not necessarily need to reproduce the entire range of the seismic properties of BLAPs. 

The bottom row in Fig.~\ref{fig:period_and_eta} shows the expected properties of the first radial overtone. The left panel shows how the ratio of the period of the first overtone to the period of the fundamental mode changes over time. Importantly, this ratio is not constant, but varies significantly when the model crosses the BLAP region, ranging from about 0.755 to 0.815. Analysis of the stability of the first radial overtone (lower right panel) suggests that BLAPs born from DELM WDs can have both the fundamental radial mode and the first radial overtone excited simultaneously provided that the initial $Z$ is high enough. In contrast, the second radial overtone is stable in all our models. Currently, only one case of a double-mode BLAP is known, most likely with two radial modes excited. This object is OGLE-BLAP-030 \citep[][their Sect.~4.2]{2024arXiv240416089P}. It shows three independent and some combination frequencies. The period ratio of the two dominant terms is equal to 0.762. This value is fully consistent with our predictions shown in Fig.~\ref{fig:period_and_eta} (bottom left panel). Although this does not prove that OGLE-BLAP-030 was created by the DELM WD merger, the possibility of using observed modes in this star to verify BLAP formation scenarios seems interesting.
\begin{figure}[!ht]
\centering
\includegraphics[width=\hsize]{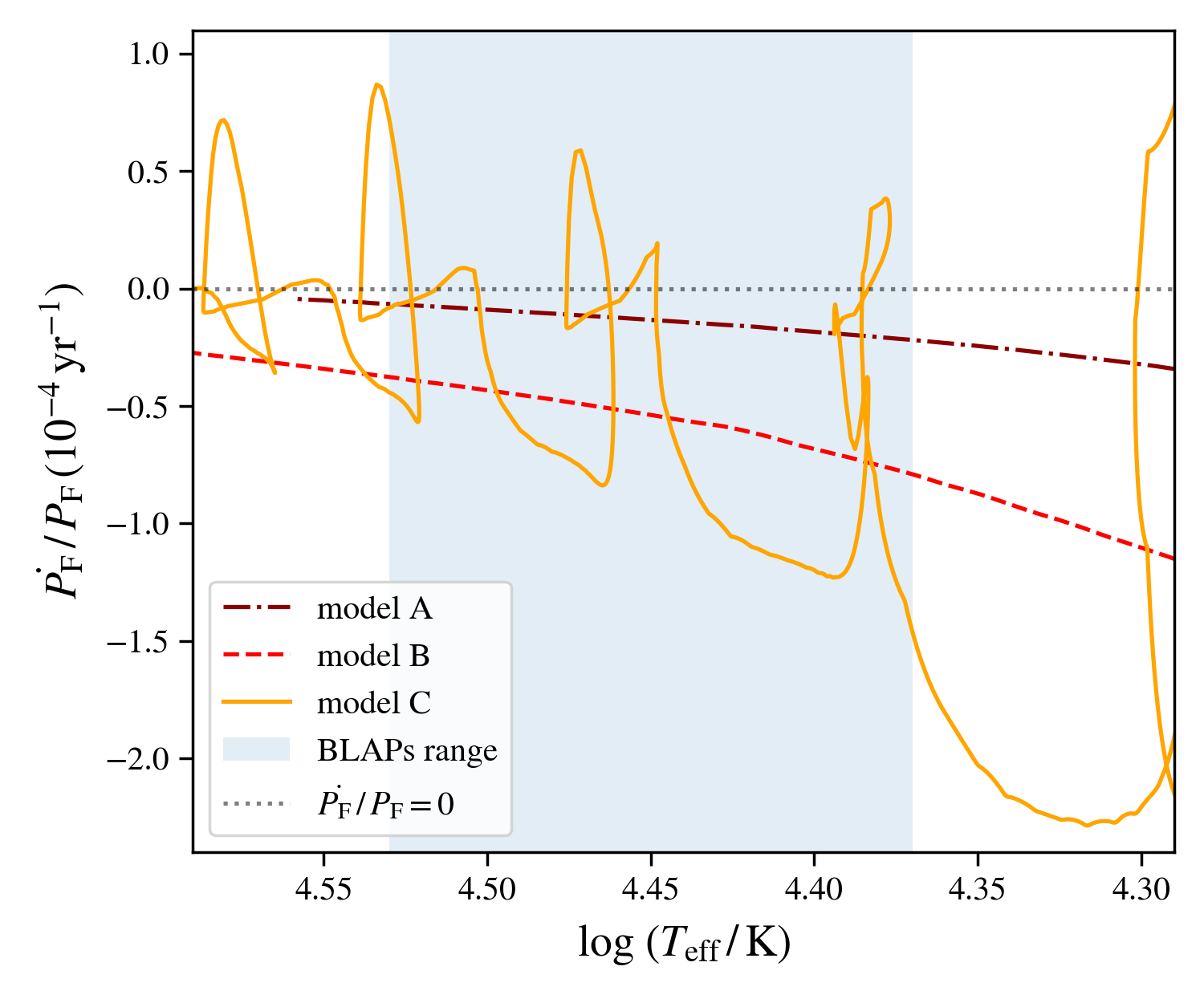}
\caption{Evolution of the rate of change of the pulsation period of the fundamental radial mode, normalised to its instantaneous value, for models A (dashed-dotted curve), B (dashed curve), and C (solid curve) with an initial $Z=0.02$ as a function of effective temperature. The shaded vertical region indicates the observed effective temperature range of the BLAPs. The dotted horizontal line indicates no change in period. The abscissa axis has been inverted to correspond to the H-R diagram; hence, time flows from right to left in the figure. Negative values on the ordinate axis indicate a shortening of the pulsation period, while positive values indicate a lengthening.}
\label{fig:pdot}
\end{figure}

\subsubsection{Rates of period changes}\label{sect:pdot}
We also investigated the rate of change of the pulsation period of the fundamental radial mode $P_{\rm F}$ over time $t$, $\dot{P_{\rm F}}\equiv{\rm d}P_{\rm F}/{\rm d}t$, which can be expected from our models when passing through the region occupied by BLAPs in the Kiel diagram. Figure~\ref{fig:pdot} shows the results of this analysis for models with initial $Z=0.02$ (for $Z=0.01$ the results are almost the same). In general, BLAPs resulting from the DELM WD mergers may show an increasing or decreasing $P_{\rm F}$ at a typical order of $\dot{P}_{\rm F}/P_{\rm F}\sim\pm10^{-5}$\,--\,$10^{-4}\,{\rm yr}^{-1}$. The direction of this change, however, depends on whether the post-merger product has managed to ignite helium when crossing the BLAP region. The models A and B only show a decrease in $P_{\rm F}$ (i.e.~negative values of $\dot{P}_{\rm F}$), as the radius of these stars decreases at this stage of evolution. Moreover, as long as helium ignition does not occur, the rate of period change is the smaller the lower is the mass of the merger product. For the model C, which represents a scenario with off-centre He-burning BLAPs, the star undergoes both a contraction and an expansion phase, which means that the pulsation period can either increase or decrease.

Although the pulsation periods of BLAPs are relatively short, which allows for an accurate analysis of the changes of $P_{\rm F}$, it may not be straightforward to determine evolutionary changes that could be compared to those in Fig.~\ref{fig:pdot}. For most BLAPs, the analysis of the period changes has been done by dividing the light curves into shorter parts and fitting a period for each part separately \citep[e.g.][]{2017NatAs...1E.166P,2024arXiv240416089P}. Such a procedure could give evolutionary period changes under the assumption that the rate of period change is constant. Unfortunately, the observed changes could be more complicated as in the case of TMTS-BLAP-01 \citep{2023NatAs...7..223L} or the two BLAPs discussed in the accompanying paper by \cite{Pigulski+2024}, and could also include periodic component due to the light travel-time effect in a binary system.

\section{Population synthesis of Galactic DELM WDs}\label{sect:pop-synth-general}
In the previous section, we showed that the post-merger product of a DELM WD can become a BLAP, provided that the total mass of the system is in the range 0.32\,--\,0.7\,M$_\sun$. However, to answer the question of how common this formation channel is and how many such BLAPs may currently exist in the Galaxy, it is necessary to refer to the binary population synthesis simulations \citep[see][for a concise summary of this technique]{2020RAA....20..161H}. Such analyses for DHeWDs (or DDs in general) have been performed many times before \citep[e.g.][to list a few]{2001A&A...365..491N,2012A&A...546A..70T,2017MNRAS.470.1894K,2019ApJ...871..148L,2020ApJ...898...71B,2021MNRAS.504.2670Y,2023ApJ...945..162T}, but never in the context of BLAPs. We therefore carry out a population synthesis of DHeWDs, focusing only on systems that are progenitors of BLAPs according to our criteria.\footnote{Although we have defined the range of masses of BLAPs originating from DELM WD mergers in Sect.~\ref{sect:mesa-general}, it cannot be excluded that binary systems consisting of an ELM WD and a HeWD (with mass higher than 0.35\,M$_\sun$) are also progenitors of BLAPs, provided that their total mass is within this range.}

\subsection{Methods}\label{sect:pop-synth-methods}
To estimate the number of BLAPs originating from merging DELM WDs that may currently exist in the Galaxy, as well as the corresponding evolutionary scenarios and associated relative probabilities, we have performed a population synthesis of binary systems based on the COSMIC code\footnote{\url{https://cosmic-popsynth.github.io/docs/}} \citep[version~3.4.10;][]{2020ApJ...898...71B}. This code is largely based on the SSE \citep{2000MNRAS.315..543H} and BSE \citep{2002MNRAS.329..897H} codes designed to track the evolution of stars and their interactions in binary systems. However, COSMIC contains many modifications compared to these two original codes and is particularly suitable to study the formation of compact binary systems. Many of the solutions in the COSMIC code have been adapted from the StarTrack binary population synthesis code \citep{2008ApJS..174..223B}.

We sampled the initial population using a multidimensional distribution of initial parameters for binary systems provided by \cite{2017ApJS..230...15M}. This was done in the COSMIC by a built-in function that iteratively adjusted the parameters of the initial population to focus only on the group of binary systems that are likely to complete their evolution at a user-specified stage. In our case, we are interested in a population that would produce DHeWDs. Our so-called fixed population contains 1.5\,$\times$\,10$^6$ of the binary systems of interest, which correspond to a total mass of all stars (both single and in binary systems) of about 1.8\,$\times$\,10$^7$\,M$_\sun$ or 1.5\,$\times$\,10$^7$ stars. We normalised the fixed population to the Galactic (`astrophysical') population based on the total mass of stars in the Milky Way, which is estimated to be 6.43\,$\times$\,10$^{10}$\,M$_\sun$ in the current epoch \citep{2011MNRAS.414.2446M}.

\begin{figure}
\centering
\includegraphics[width=\hsize]{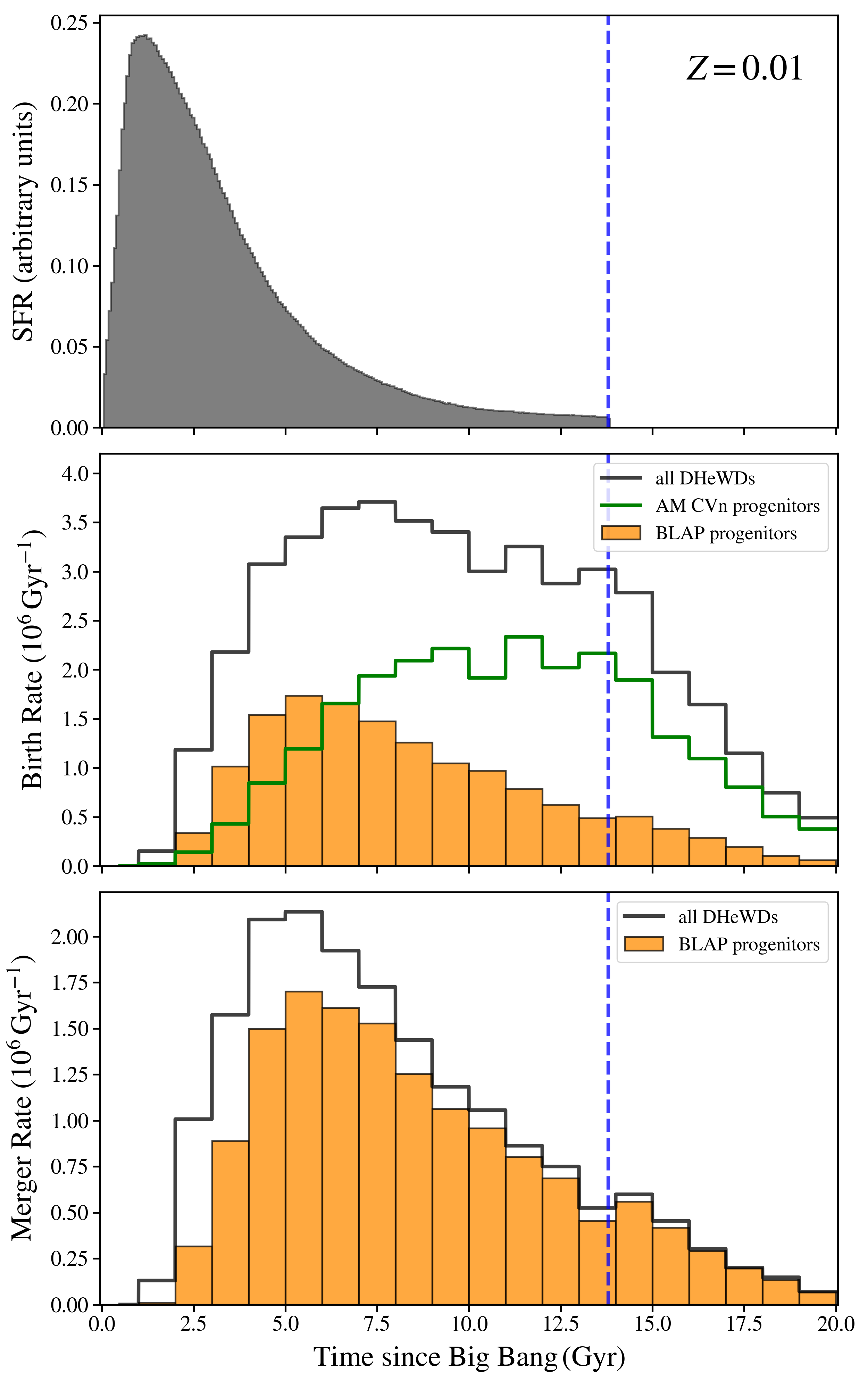}
\caption{Summary of our population synthesis of DHeWDs performed using the COSMIC code for metallicity $Z=0.01$. The upper panel shows the SFH used in the simulation. The middle panel shows the birth rate of all DHeWDs (black histogram), as well as those among them that will become AM~CVn-type systems (green histogram) or BLAPs (orange histogram). The lower panel presents the merger rate of all DHeWDs with the corresponding mass ratio (grey histogram) and those systems that will become BLAPs (orange histogram). The blue vertical line marks the age of the Universe.}
\label{fig:merger_rates_histograms}
\end{figure}

The fixed population was generated using a custom star formation history (SFH) assuming that the current age of the Universe is 13.8\,Gyr \citep{2020A&A...641A...6P}. We have adopted the star formation rate (SFR) profile from \cite{2016A&A...588A..21R} resulting from the chemo-dynamical modelling of a Milky Way-like galaxy performed with proprietary GEAR code \citep{2012A&A...538A..82R}. In addition to the separate treatment of the stellar, interstellar medium, and dark matter components, the model also allows for the direct tracking of [Fe/H] and [Mg/Fe] abundances by solving the chemical evolution equations explicitly. It was shown to accurately reproduce the observed [Fe/H]–[Mg/Fe] scatter as obtained from high-resolution spectroscopy of stars in the Milky Way and dwarf spheroidal galaxies in the Local Group. Since \cite{2016A&A...588A..21R} do not provide a digital version of their SFR function, we digitised their Fig.\,17, averaged, and smoothed all three curves shown therein. The result is shown in Fig.~\ref{fig:merger_rates_histograms} (upper panel). The SFH we used shows a maximum around 1.5\,Gyr after the Big Bang and an almost exponential decrease in time. We also assumed that the adopted multi-dimensional distribution of initial parameters for binary systems is universal and time-independent.

Since most of the stars in our sample formed $\sim$1.5\,Gyr after the Big Bang (according to the adopted SFH), we have assumed that the entire population is characterised by a metallicity ${\rm [Fe/H]}=-0.3$ \citep{2023NatAs...7..951L}. Assuming that the solar metallicity is $Z_\sun=0.02$ \citep{2016ApJ...816...13V}, this corresponds to $Z=0.01$. However, we also repeated our calculations for $Z=Z_\sun$ to investigate the impact of this parameter on our results, as we describe below in the text. The wind mass-loss rate on the RGB was calculated using a Reimers-like wind with a dimensionless scaling factor $\eta_{\rm R}=0.48$ \citep{2015MNRAS.448..502M}. For the CE phase, we used the canonical value of the efficiency of orbital energy transfer to the envelope, $\alpha_{\rm CE}=1$. To distinguish between stable and unstable mass transfer depending on the evolutionary stage of the components, we adopted the set of critical mass ratios from option number `3' in the COSMIC (\texttt{qcflag=3} in the \texttt{bse} section of input parameters), which is recommended by the documentation of this code for the synthesis of DDs. The rest of the free parameters and `switches' in the COSMIC code remained mostly default. The code we used to generate the fixed population is available on Zenodo\footnote{\url{https://doi.org/10.5281/zenodo.13863170}}.

\subsection{Birth and merger rates}\label{sect:pop-synth-rates}
Among the 1.5\,$\times$\,10$^6$ binary systems whose evolution we traced with COSMIC, we identified those that led to the formation of DHeWD. We considered two possibilities for their further evolution. Firstly, the system could undergo a merger during the CE phase or coalesce as a result of gravitational radiation. In the latter scenario, we assumed that the binary orbit is circular and shrinks according to the formalism provided by \cite{1964PhRv..136.1224P}. In the case of mergers, we assumed that they occur when the distance between components in a circular orbit is equal to the sum of their radii. Secondly, a DHeWD system can become an AM~CVn-type binary if the mass ratio of the donor to accretor is less than $q_{\rm crit}\approx2/3$ \citep{2007ApJ...670.1314M}, leading to a stable mass transfer between the components.\footnote{In our study we assume $q_{\rm crit}=0.625$, according to the choice of \texttt{qcflag} in the COSMIC input file.} Systems of this kind follow different evolutionary paths than DHeWD mergers and are not investigated in this article.

According to the results presented in Sect.~\ref{sect:mesa-results}, we identified DHeWDs that can be the progenitors of BLAPs as those DHeWDs with a total mass not exceeding 0.7\,M$_\sun$ and a mass ratio that allows for a dynamic merger instead of forming an AM~CVn-type binary system. We did not need to consider the lower mass limit for BLAPs, as all DHeWDs in our population have total masses above 0.32\,M$_\sun$. A summary of our population synthesis is shown in Fig.~\ref{fig:merger_rates_histograms}. The middle panel shows the predicted birth rates of the different types of systems. DHeWDs are mostly born between 5 and 15\,Gyr after the Big Bang. In this time interval, such systems in the whole Galaxy are formed at a rate of about 3\,$\times$\,10$^6$\,Gyr$^{-1}$. However, those that can become BLAPs are mainly born at around 6\,Gyr after the Big Bang at a rate of about 1.5\,$\times$\,10$^6$\,Gyr$^{-1}$. In contrast, the progenitors of AM~CVn-type systems form with the highest rate 7 to 15\,Gyr after the Big Bang.

The lower panel of Fig.~\ref{fig:merger_rates_histograms} shows the merger rates for all DHeWDs that may have undergone unstable mass transfer and for the progenitors of BLAPs. These rates peak at about 6\,Gyr after the Big Bang. Our simulations suggest that the vast majority of the presently merging DHeWDs are progenitors of BLAPs, with approximately 600\,000 stars of this type per Gyr formed in the entire Galaxy. Assuming that the merger product becomes a BLAP shortly after the merger completes (which is a realistic assumption according to Sect.~\ref{sect:mesa-results}) and that the time of the passage through the region in which fundamental radial mode is unstable lasts up to $\sim$70\,000 years (Figs.~\ref{fig:HR_all_models} and \ref{fig:period_and_eta}), it can be estimated that there should currently be around 40 BLAPs in the Galaxy that are descendants of DHeWD (including DELM WDs) mergers. This means that in the BLAP formation scenario we studied, their rarity is the result of relatively fast evolution rather than the small number of DHeWD mergers.

\begin{figure}
\centering
\includegraphics[width=\hsize]{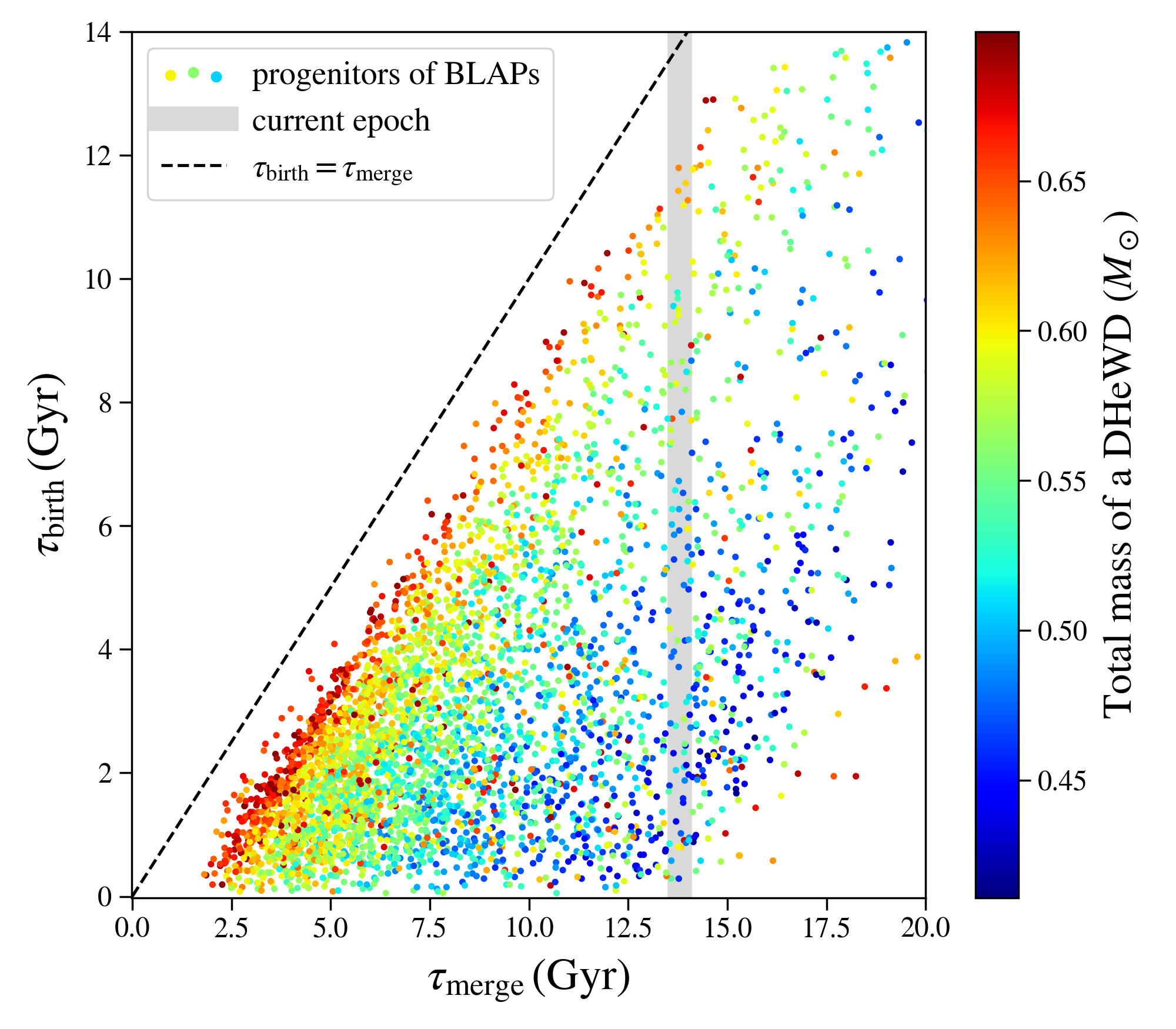}
\caption{Age of the formation of the initial binary system, $\tau_{\rm birth}$ as a function of the age of the corresponding DHeWD merger, $\tau_{\rm merge}$. Both ages are calculated from the Big Bang. The colour scale corresponds to the total mass of the DHeWD. The dashed line represents the 1:1 relationship, while the vertical shaded bar indicates the current epoch.}
\label{fig:tbirth_tmerge}
\end{figure}

We also investigate from which epochs of Galactic evolution the progenitors of the currently observed BLAPs might originate. To do so, we plotted the initial birth age of the system against the age of its DHeWD merger. The results are shown in Fig.~\ref{fig:tbirth_tmerge} and reveal a wide range of possibilities. It is clear that the progenitors of the present BLAPs, originating from the DHeWDs, could have formed practically at the beginning of the formation of the Galaxy. They are, on average, less massive than those whose initial systems only formed about 2.5\,Gyr ago. Although our simulation corresponds to $Z=0.01$, Fig.~\ref{fig:tbirth_tmerge} suggests that BLAPs formed by DELM WD mergers may exhibit a very wide range of metallicities, reflecting the chemical evolution of the Galaxy. Consequently, it is to be expected that not all products of DHeWD mergers will pulsate like BLAPs when passing through their region of occurrence in the H-R diagram. This may be because the content of iron-group elements, which are responsible for driving radial pulsations in BLAPs, does not change significantly in the post-merger product compared to the initial content in the MS phase. Consequently, those mergers whose progenitors formed early in the evolution of the Galaxy, in low-metallicity environment, may not pulsate even though they have an internal structure similar to BLAPs.

The population synthesis we performed suffers from many uncertainties that are difficult to quantify. By changing the metallicity of the population to $Z=0.02$ and leaving all other parameters unchanged, we verified that both the birth and merger rates of all DHeWDs decrease (Fig.~\ref{fig:merger_rates_histograms_zsun}). Apart from the obvious limitations and approximate nature of the SSE and BSE codes, the main sources of uncertainty are (i) the normalisation of the fixed population to the astrophysical one, (ii) the assumption that the initial distribution of binary system parameters does not depend on the age of the Galaxy, and (iii) treating the Galaxy as a homogeneous stellar population without taking into account the separate chemical and dynamical evolution of stars in different components of the Galaxy. One can realistically expect that the uncertainties associated with the effects of our population synthesis can accumulate to an order of magnitude. Thus, it can be summarised that the estimated number of BLAPs in the Galaxy, formed in the merger scenario analysed here, could range from a few to several hundred.

\begin{figure}
\centering
\includegraphics[width=\hsize]{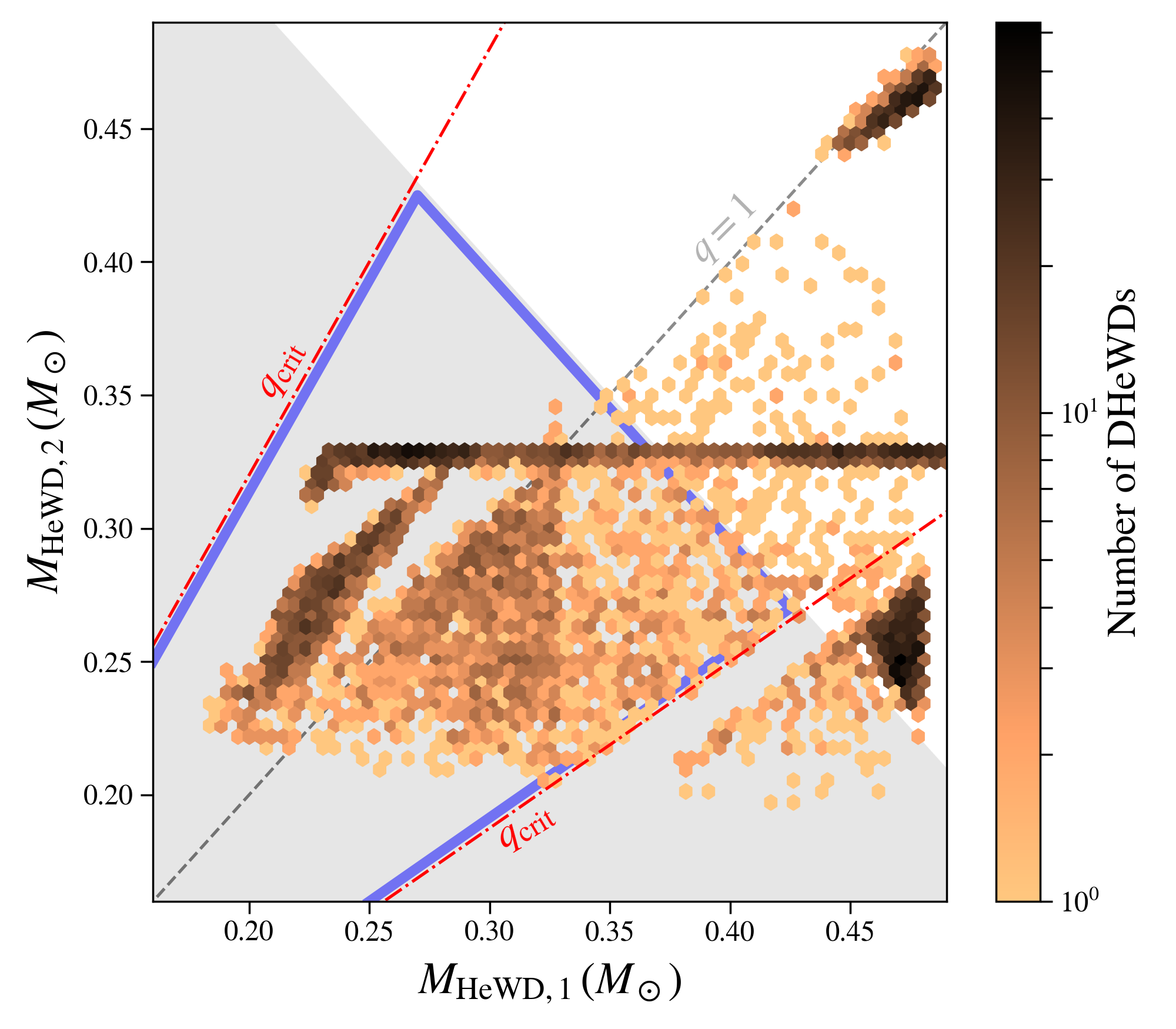}
\caption{Number of DHeWDs (colour-coded) from our binary population synthesis on the mass-mass plane. Index `1' corresponds to the component that was more massive on the MS, and index `2' corresponds to the less massive component. The grey dashed line corresponds to equal component masses. The red dotted-dashed lines surround the area where the mass ratio of the components allows for a merger, $q>q_{\rm crit}=0.625$. The grey shaded area corresponds to the sum of the masses of the components smaller than 0.7\,M$_\sun$. The blue polygon encloses the region with systems that may be progenitors of BLAPs. }
\label{fig:DHeWDs_m1m2_hexbin}
\end{figure}

\subsection{Evolutionary scenarios}\label{sect:pop-synth-scenarios}
Figure~\ref{fig:DHeWDs_m1m2_hexbin} shows our sample of DHeWDs on the plane of the masses of their components. There are at least several groups of systems on this plane that correspond to different evolutionary scenarios. The diagram is populated on both sides of the grey dashed line, which denotes equal component masses (i.e.~the ratio of component masses $q=1$). Thus, there are post-RLOF systems with $q>1$ in which the more massive HeWD originates from the initially less massive secondary component. The systems of interest from the point of view of BLAP formation are located in the region bounded by the pair of red dash-dotted lines. This is the area where the mass ratio allows for a dynamic merger. At the same time, the DHeWDs from which the BLAPs originate must lie within the grey-shaded area where the total mass of the system is smaller than 0.7\,M$_\sun$. A series of panels analogous to Fig.~\ref{fig:DHeWDs_m1m2_hexbin}, but with different parameters of DHeWDs coded with colours, can be found in Fig.~\ref{fig:DHeWDs_m1m2_planes}.

\begin{figure*}
\centering
\includegraphics[width=\hsize]{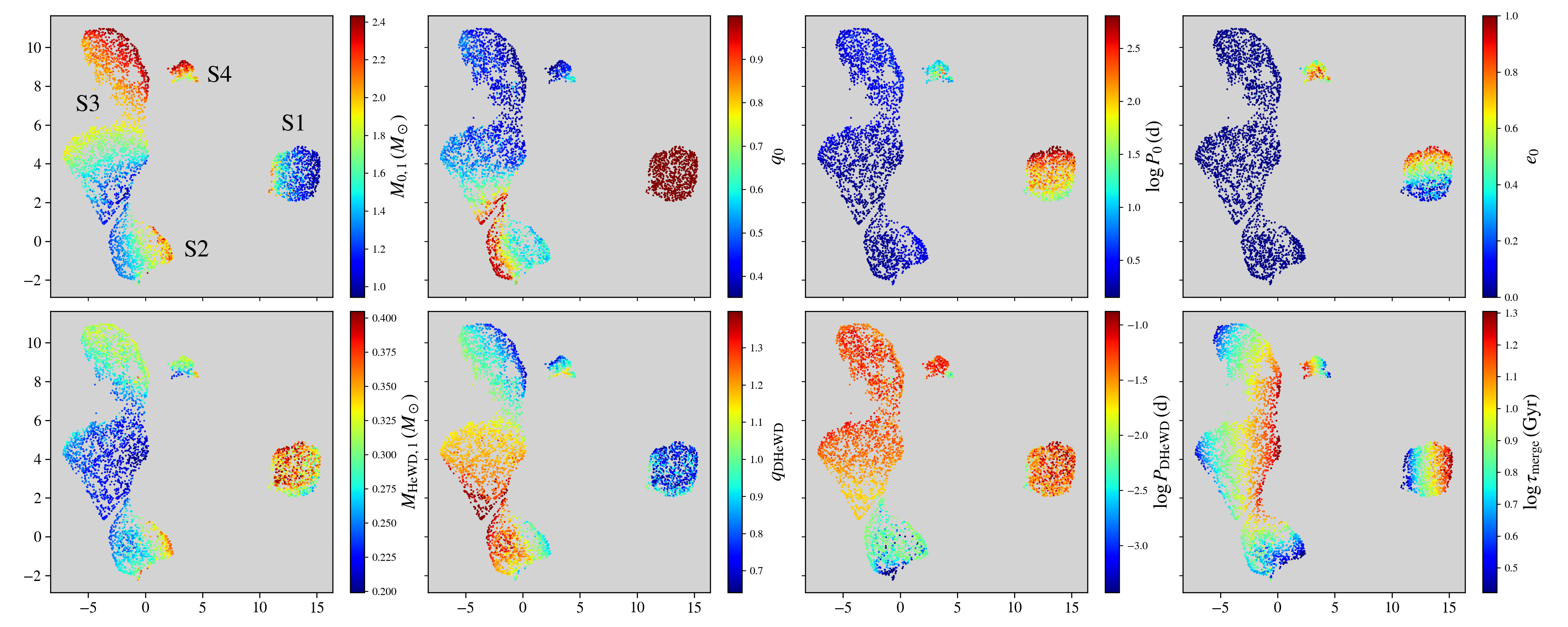}
\caption{Two-dimensional UMAP projection of the manifold spanned by the parameters of our DHeWD sample, whose merger can result in the formation of BLAPs. One can see the presence of four distinct evolutionary scenarios, which are labelled in the upper left panel. The colour scale on each panel corresponds to a different parameter. Starting from left to right these parameters are: upper row: initial mass of the primary component ($M_{0,1}$), which is the more massive one on the MS, the initial mass ratio ($q_0$), the logarithm of the initial orbital period ($\log P_0$) and initial orbital eccentricity ($e_0$); lower row: the mass of the HeWD formed from the primary component ($M_{\rm HeWD,1}$), the mass ratio in the DHeWD system ($q_{\rm DHeWD}$), the logarithm of the orbital period of the DHeWD ($\log P_{\rm DHeWD}$) and the logarithm of the age at which the DHeWD merger occurs ($\log \tau_{\rm merger}$). We emphasize that the dimensions on both axes are not physically meaningful and their values are the result of a non-linear UMAP projection.}
\label{fig:umap_blaps_only}
\end{figure*}

\begin{table*}
\caption{\label{tab:pop-synth}Summary of four evolutionary scenarios leading to the formation of BLAPs by merging DELM WDs.}
\centering
\begin{tabular}{crlccccccc}
\hline\hline
\noalign{\smallskip}
Scenario&Fraction&Evolution&$M_{0,1}$&$q_0$&$\log P_0$&$e_0$&$M_{\rm HeWD,1}$&$q_{\rm DHeWD}$&$\log P_{\rm DHeWD}$\\
&(\%)&&$({\rm M}_\sun)$&&(d)&&$({\rm M}_\sun)$&&(d)\\
\noalign{\smallskip}
\hline
\noalign{\medskip}
S1&20.6&double-core&(1.0;2.2)&$\sim1$&(1.5;2.6)&(0;0.9)&(0.25;0.40)&(0.8;1.0)&($-$2.3;$-$1.0)\\
&&CE&&&&&&&\\
\noalign{\smallskip}
S2&23.2&RLOF\,+\,CE&(1.3;2.2)&(0.55;0.95)&(0.3;0.6)&0&(0.23;0.37)&(0.95;1.3$^\ast$)&($-$3.2;$-$2.2)\\
\noalign{\smallskip}
S3&51.4&RLOF\,+\,CE&(1.0;2.4)&(0.35;0.95)&(0.3;0.7)&0&(0.20;0.33)&(0.7;1.4)&($-$1.7;$-$1.2)\\
\noalign{\smallskip}
S4&4.8&RLOF\,+\,CE&(1.8;2.4)&(0.40;0.60)&(0.8;2.5)&(0.2;0.95)&(0.22;0.35)&(0.7;1.2)&($-$2.2;$-$1.1)\\
\noalign{\smallskip}
\hline
\end{tabular}
\tablefoot{The symbols of parameters in the first row are identical to those presented and described in the caption to Fig.~\ref{fig:umap_blaps_only}. $^\ast$ The secondary component of a DHeWD has a nearly constant mass of about $0.33\,{\rm M}_\sun$.}
\end{table*}

In order to distinguish between different evolutionary scenarios leading to the formation of BLAPs through mergers of DHeWDs and thus classify the sample shown in Fig.~\ref{fig:DHeWDs_m1m2_hexbin}, we utilised a dimensionality reduction technique called Uniform Manifold Approximation and Projection\footnote{\url{https://umap-learn.readthedocs.io/en/latest/}} \citep[UMAP,][]{2018arXiv180203426M}. We applied this method to an 11-dimensional space containing the basic properties of the simulated DHeWDs and projected the original manifold of our sample onto the 2-dimensional (2D) plane. The analysis was performed only for those DHeWDs that, according to our criteria, could lead to the formation of BLAPs. The results are summarised in Fig.~\ref{fig:umap_blaps_only}. Each panel represents the same 2D projection of the sample, colour-coded according to the different key parameters of DHeWDs. It can be seen that there are four well-defined and isolated groups of points, corresponding to four distinct evolutionary scenarios. For simplicity, we have labelled these scenarios as S1, S2, S3, and S4 in the upper left panel of Fig.~\ref{fig:umap_blaps_only}. We describe their key features below.
\begin{itemize}
\item [S1.] The first scenario corresponds to the evolution of binary systems with an initial mass ratio very close to unity. Initially, the system is characterised by a relatively long orbital period (up to about 300 days), and its orbit can be highly eccentric. Due to the equal masses of the components, both begin to leave the MS at a similar time. As their radii rapidly grow during the evolution to the RGB, the orbit quickly circularises, leading to the `double-core' common envelope (CE) phase \citep[e.g.][]{1995ApJ...440..270B}. Matter to the envelope is transferred simultaneously from both components. In this way, both HeWDs are born at virtually the same time.
\item [S2.] The initial binary system has a close configuration because the orbital period is only about 2 days long. As the primary component begins to cross the Hertzsprung gap, it easily fills its Roche lobe and stable mass transfer to the secondary component begins. The primary component sheds its hydrogen-rich envelope and becomes an ELM WD, while the binary system behaves at this stage as the EL~CVn-type system \citep[e.g.][]{2017MNRAS.467.1874C}. When the secondary component reaches the RGB, a CE phase occurs, leaving the tightest configuration of all four scenarios, with a DHeWD with an orbital period of about 10 minutes. A characteristic feature of the S2 scenario is also that the secondary component eventually becomes a more massive HeWD than the remnant of the primary component, and its mass is almost constant and equal to $\sim$0.33\,M$_\odot$ (horizontal feature in Fig.~\ref{fig:DHeWDs_m1m2_hexbin}).
\item [S3.] In this scenario, the progenitors of BLAPs are relatively close circular binary systems with a wide range of initial component masses. As in the S2 scenario, the system first undergoes the Roche-lobe overflow (RLOF) phase and then a CE phase. However, in contrast to the S2 scenario, the interplay between the initial conditions leads to the formation of DHeWDs with much longer orbital periods of up to 100 minutes. The characteristic `split' of this group into two parts on the UMAP plane is a consequence of the distinction between systems that have or have not reversed their mass ratio during the DHeWD stage.
\item [S4.] In terms of outcome and evolutionary channel, this scenario is almost the same as S3. However, the key difference between the two concerns the initial population. In contrast to the relatively close and circular initial orbits in the S3 scenario, those in S4 are highly eccentric and have relatively long orbital periods of up to several hundred days. The implementation of this scenario also requires significant mass discrepancy between the MS components. Due to pseudosynchronisation and significant eccentricity, both components spin up significantly during the MS phase. When the primary component evolves towards the RGB, the system undergoes rapid circularisation and a reduction of the semi-major axis due to strong tides. The further evolution is similar to that of S3.
\end{itemize}
A detailed summary of the characteristics of all four scenarios can be found in Table~\ref{tab:pop-synth}. We have included typical parameter ranges and the percentage of the occurrence of each scenario. The latter statistic is integrated over the time of the Galaxy evolution, thus referring to the total number of BLAPs resulting from HeWD mergers in all epochs. The most likely scenario is the S3 scenario.

\section{Summary and conclusions}\label{sect:summary}
In our study, we tested whether DELM WDs or DHeWDs in general could be progenitors of BLAPs (Sects.~\ref{sect:introduction} and \ref{sect:binary-wds}). To answer this question, we modelled the mergers of DELM WDs and the evolution of their products (Sect.~\ref{sect:mesa-general}). We also analysed the seismic properties of these models (Sects.~\ref{sect:seismic-properties} and \ref{sect:pdot}). Finally, we estimated the number of Galactic BLAPs that could originate from this evolutionary channel by means of a dedicated binary population synthesis with the COSMIC code (Sect.~\ref{sect:pop-synth-general}). The main results of our analysis are as follows:
\begin{itemize}
\item[$\bullet$]We find that BLAPs can be formed from a merger of a DELM WD system with a total mass in the range 0.32\,--\,0.7\,M$_\odot$. This mass range is virtually the same for models with initial metallicity of progenitors $Z=0.01$ and $Z=0.02$.
\item[$\bullet$]The post-merger product crosses the BLAP region on the Kiel diagram only four to ten thousand years after the coalescence of the components. The crossing time of the merger product through the BLAP region ranges from approximately 20 to 70 thousand years.
\item[$\bullet$]Mergers formed in this way either have no thermonuclear reactions anywhere in their interiors when crossing the BLAP region (models A and B) or undergo off-centre helium burning (model C). The models B and C eventually become hybrid He/CO WDs.
\item[$\bullet$]Seismic analysis of our models reproduces pretty well the region in which BLAPs are observed. In addition to the instability of the fundamental radial mode, we also found the first radial overtone for models with a higher initial metallicity to be unstable. Their period ratio ranges between 0.755 and 0.815.
\item[$\bullet$]The resulting rates of period changes for the fundamental radial mode, $\dot{P}_{\rm F}/P_{\rm F}$, are negative for the models A and B, positive and negative for model C, and can reach $\pm\,10^{-4}\,{\rm yr}^{-1}$.
\item[$\bullet$]Population synthesis showed that up to a few hundred BLAPs originating from the proposed merger scenario can exist at the present day in the Galaxy.
\item[$\bullet$]The DELM WD systems that are progenitors of BLAPs are born in four evolutionary scenarios, with the most likely scenario being the system undergoing a stable mass transfer between the components, followed by the CE phase.
\end{itemize}

Based on the results obtained in our study, we can draw the following conclusions. Our simulations suggest that BLAPs, which are descendants of DELM WD mergers, can pulsate simultaneously in the fundamental radial mode and the first radial overtone provided that the initial metallicity of their progenitors is $Z\gtrsim0.01$. This theoretical result differs from predictions in previous studies on the origin of BLAPs. In the paper reporting the discovery of BLAPs, \cite{2017NatAs...1E.166P} indicated that both radial modes in their models are stable, although the first overtone is more difficult to excite than the fundamental mode. Furthermore, \cite{2018MNRAS.477L..30R} reported that in their models, the first overtone is stable despite synthetically increasing metallicity in the envelope to $Z=0.05$. Recently, \cite{2024arXiv240816912J} showed that it is possible to obtain seismic models of BLAPs with the first overtone excited using the MESA-RSP code \citep{2008AcA....58..193S,Paxton2019}. However, the methods used and the assumptions made by the authors can be considered questionable in the context of BLAPs. Firstly, the authors adopted an artificially increased metallicity $Z=0.05$ while retaining the canonical value of $X=0.7$, which is at odds with the significant overabundance of helium observed in BLAPs. Secondly, the MESA-RSP code performs the seismic analysis on a simplified model of an isolated and chemically homogeneous outer envelope, which is constructed independently of the evolutionary scenario and the actual internal structure of the whole BLAP. For these reasons, it is difficult to attribute the models presented by \cite{2024arXiv240816912J} to a specific evolutionary history, whereas previous studies on the origin of BLAPs have shown that there is certainly a link between their seismic properties and origin. Unfortunately, there is no information on the stability of the first overtone in other papers on BLAPs. This includes the study by \cite{2023ApJ...959...24Z}, which investigated the possibility of BLAP formation from MS-WD mergers. It is therefore unclear whether the potential observation of both radial modes in a BLAP (OGLE-BLAP-030 is an example) is unambiguous evidence that it originated from the coalescence of the components of a DELM WD system.

We would also like to emphasize that both radial modes in our models are excited without needing to segregate the elements in the outer layers. This contrasts with the conclusions presented by \cite{2020MNRAS.492..232B}, who argued that the excitation of radial oscillations in BLAPs is only possible with radiative levitation and gravitational settling of elements (leading to the iron accumulation in the $Z$-bump region). There are several factors leading to such different results. First of all, \cite{2020MNRAS.492..232B} analysed a completely different evolutionary scenario compared to ours. The authors considered mass stripping of an RGB star via stable or unstable mass transfer. Consequently, their BLAPs still contain a low-mass hydrogen envelope, which significantly affects the seismic properties. The hydrogen envelope provides an additional source of opacity and makes the $Z$-bump less pronounced, hence the need for atomic diffusion to enhance the $Z$-bump and drive the pulsations. Secondly, the models presented by \cite{2020MNRAS.492..232B} do not account for rotational effects that counteract diffusion of elements. This is particularly important in our scenario because the merger product rotates relatively fast and is characterised by intense rotational mixing in the outer layers. Thirdly, some of the models proposed by the aforementioned authors can undergo shell hydrogen-burning, making their evolution time through the BLAP region long enough for segregation processes to take place. In contrast, our models are relatively fresh post-merger products, and they evolve much faster through the BLAP region. Even without rotational mixing, it is unlikely that there is enough time for radiative levitation and gravitational settling to take effect.

The most optimistic observation-based estimate of the total number of BLAPs present in the Galaxy provides a number around 28\,000 \citep{2020ApJ...903..100M}. Up to several hundred BLAPs originating from merging DELM WDs seem to represent a small, albeit non-negligible, fraction of this number. \cite{2021MNRAS.507..621B}, on the other hand, estimate that there should be about 12\,000 BLAPs in the entire Galaxy that are He-core, shell H-burning pre-WDs. Furthermore, \cite{2022A&A...668A.112X} claim that between 3\,500 and 70\,000 Galactic BLAPs can be expected that are shell He-burning sdB-type hot subdwarfs. The number of BLAPs that have survived as companions of type Ia supernovae may currently range between 750 and 7\,500 \citep{2020ApJ...903..100M}. Given the number of difficult-to-verify assumptions needed to derive these estimates, it should be noted that they are highly uncertain. This is evidenced by the fact that the sum of these estimates far exceeds the maximum number of Galactic BLAPs estimated from the observed sample.

Simulations of merging DDs suggest that a steep rotational velocity gradient during the dynamic phase of merger should lead to a significant enhancement of the magnetic field in the post-merger product \citep[e.g.][]{2012ApJ...749...25G,2013ApJ...773..136J,2014MNRAS.438..169B,2024arXiv240702566P}. On the other hand, the CE phase may also result in an enhancement of the magnetic field of the DELM WD component(s) \citep[e.g.][]{2008MNRAS.387..897T,2016MNRAS.462L.121O}, which may later become BLAPs. Therefore, it seems that BLAPs originating from mergers should be characterised by strong magnetic fields that are measurable by spectropolarimetric techniques \citep{2015psps.book.....K}. Moreover, the radial modes of such `magnetic BLAPs' may orient their pulsation axis to coincide with the magnetic axis instead of the rotational one. Consequently, their pulsations can be described by the so-called oblique rotator model in which roAp stars are explained \citep{1982MNRAS.200..807K,2011A&A...536A..73B}. In the presence of a strong magnetic field, radial modes no longer maintain spherical symmetry and gain components of higher-degree spherical harmonics \citep[e.g.][]{2000ApJ...531L.143S}. As a result, the frequency spectra computed for the light curves of such objects show multiplets with uniform frequency spacing. In the accompanying paper \citep{Pigulski+2024}, we present two such examples of BLAPs. Their unusual light curves can be explained based on the oblique rotator model, and they are therefore natural candidates for magnetic BLAPs that could have formed as a result of DELM WD mergers.

\begin{acknowledgements}
We are grateful to the anonymous referee for carefully reading the manuscript and providing us with many useful comments and suggestions that helped us significantly improve our paper. This work was supported by the National Science Centre, Poland, grant no.~2022/45/B/ST9/03862. This research made use of \texttt{py\_mesa\_reader} \citep{2017zndo....826958W}, PyGYRE (\url{https://pygyre.readthedocs.io/en/stable/}), NumPy \citep{harris2020array}, SciPy \citep{2020SciPy-NMeth} and Matplotlib \citep{Hunter:2007}. This research has made use of the VizieR catalogue access tool, CDS, Strasbourg, France \citep{10.26093/cds/vizier}. The original description 
of the VizieR service was published in \citet{vizier2000}. This research has made use of NASA's Astrophysics Data System Bibliographic Services.
\end{acknowledgements}

\bibliographystyle{aa}
\bibliography{bib}

\begin{appendix}

\section{MESA micro- and macrophysics data}\label{appendix:mesa}
Our work uses the MESA stellar evolution code, which incorporates a vast compilation of knowledge, mainly from micro- and macrophysic, collected by many authors. The \texttt{MESAeos} module is a mixture of OPAL \citep{Rogers2002}, SCVH \citep{Saumon1995}, FreeEOS \citep{Irwin2004}, HELM \citep{Timmes2000}, PC \citep{Potekhin2010}, and Skye \citep{Jermyn2021} equations of state. Radiative opacities are taken primarily from OPAL \citep{Iglesias1993, Iglesias1996}, with low-temperature data from \citet{Ferguson2005} and the high-temperature, Compton-scattering dominated regime by \citet{Poutanen2017}.  Electron conduction opacities are from \citet{Cassisi2007} and \citet{Blouin2020}. Nuclear reaction rates are from JINA REACLIB \citep{Cyburt2010} and NACRE \citep{Angulo1999}, and additional tabulated weak reaction rates are from \citet{Fuller1985}, \cite{Oda1994}, and \cite{Langanke2000}. Screening is included via the prescription of \citet{Chugunov2007}. Thermal neutrino loss rates are taken from \citet{Itoh1996}.

\section{Additional figures}\label{appendix:figures}
\begin{figure}[h]
\centering
\includegraphics[width=0.95\hsize]{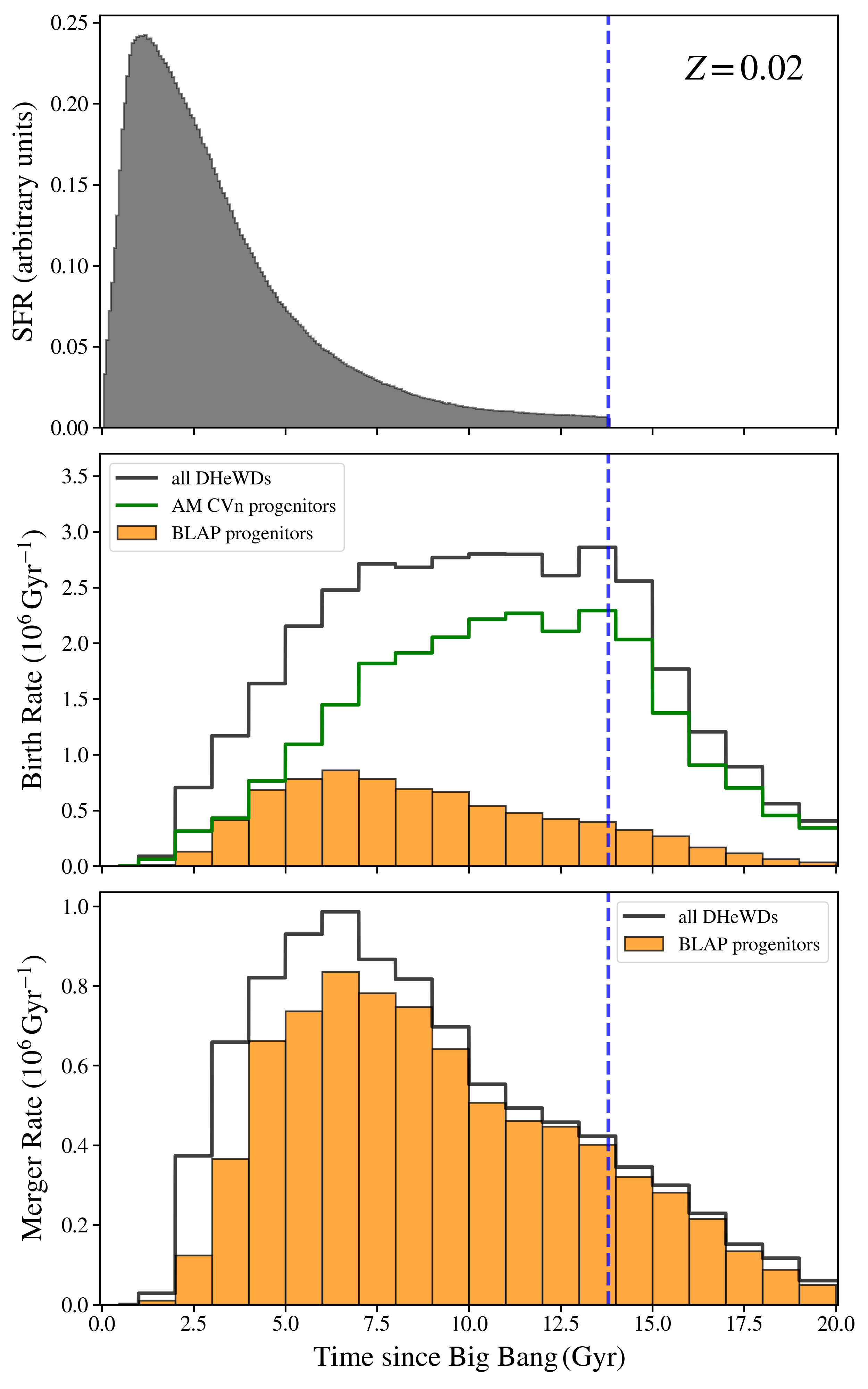}
\caption{Same as Fig.~\ref{fig:merger_rates_histograms} but for $Z=Z_\sun=0.02$. We note the smaller birth and merger rates for the progenitors of BLAPs compared to the $Z=0.01$ case.}
\label{fig:merger_rates_histograms_zsun}
\end{figure}

\begin{figure}[h]
\centering
\includegraphics[width=\hsize]{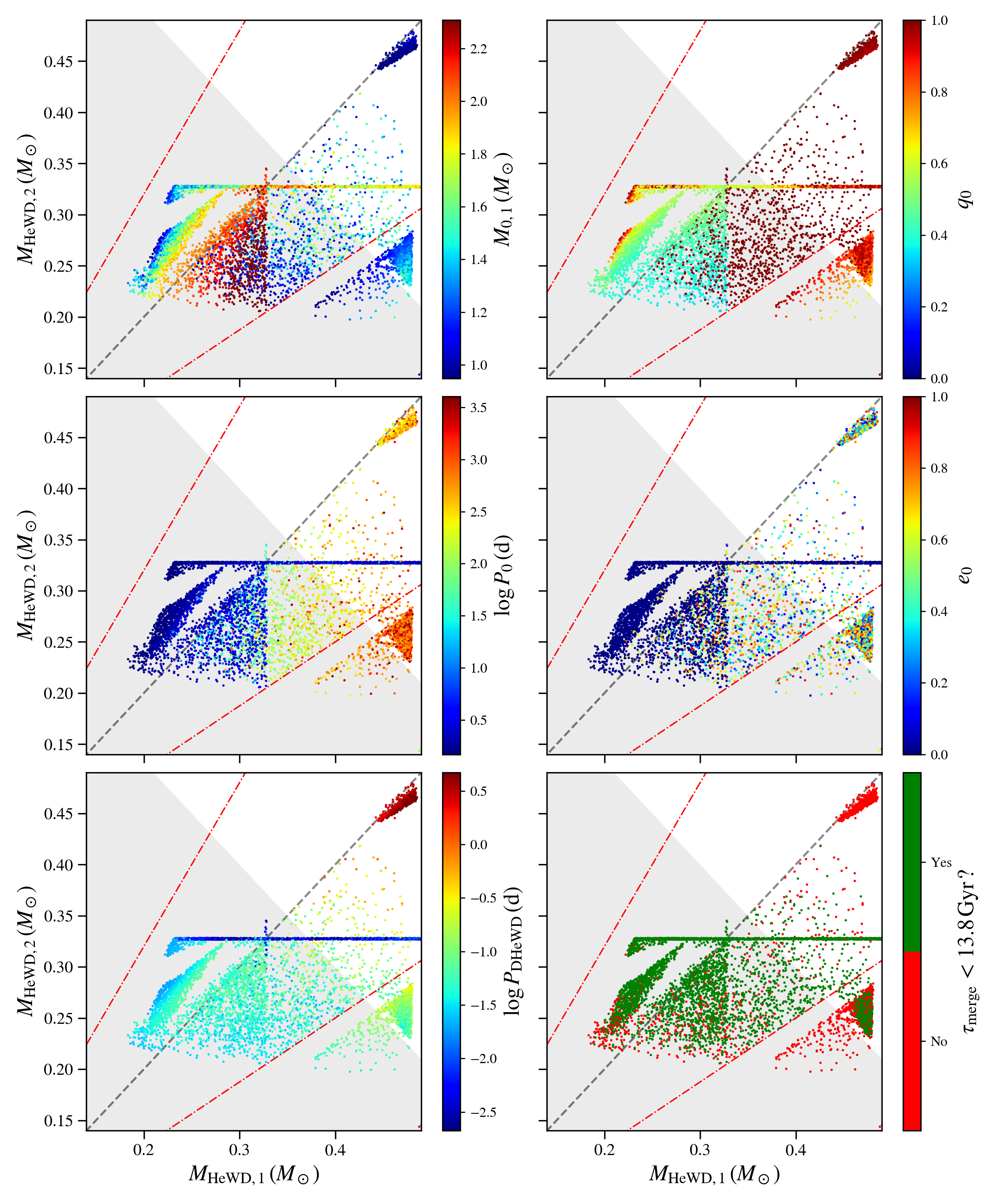}
\caption{Set of panels analogous to Fig.~\ref{fig:DHeWDs_m1m2_hexbin} but showing each generated DHeWD as a colour-coded point. The parameter symbols indicated on the colour bars have the same meaning as in Fig.~\ref{fig:umap_blaps_only}. The bottom right panel shows the distinction between systems that have already merged to the present time.}
\label{fig:DHeWDs_m1m2_planes}
\end{figure}

\end{appendix}
\label{LastPage}

\end{document}